\documentclass[sigconf,balance=true,hidelinks,table]{acmart}

\usepackage{hyperref}
\usepackage{amsmath,amsfonts}
\usepackage{algorithmic}
\usepackage[ruled,linesnumbered,noend]{algorithm2e}
\usepackage{xcolor}
\usepackage{svg}
\usepackage{graphicx}
\usepackage{textcomp}
\usepackage{xcolor}
\usepackage{nccmath}
\usepackage{environ}
\usepackage{tcolorbox}
\usepackage{soul}
\usepackage{booktabs}
\usepackage{multirow}
\usepackage{array}
\usepackage{float}
\usepackage{makecell}
\usepackage{flushend}
\usepackage{enumitem}
\usepackage{hhline}
\usepackage{comment}
\usepackage{epstopdf} 
\usepackage{caption}
\usepackage{subcaption}

\usepackage{xcolor}
\usepackage{listings}

\usepackage{scalerel}

\NewEnviron{myequation}{%
\begin{equation}
\scalebox{1}{$\BODY$}
\end{equation}
}

\newcommand{\fig}{Fig.}
\newcommand{\tab}{Table}
\newcommand{\bigvul}{Big-Vul}
\newcommand{\devign}{Devign}
\newcommand{\linevul}{LineVul}

\newcommand{\vszz}{V-SZZ}

\newcolumntype{P}[1]{>{\centering\arraybackslash}p{#1}}
\newcolumntype{R}[1]{>{\raggedleft\arraybackslash}p{#1}}

\newcommand{\code}[1]{\textnormal{\texttt{#1}}}

\acmConference[MSR 2024]{21st International Conference on Mining Software Repositories}{April 2024}{Lisbon, Portugal}

\begin{document}

\title{Are Latent Vulnerabilities Hidden Gems for\\Software Vulnerability Prediction? An Empirical Study}

\author{Triet Huynh Minh Le}
\affiliation{\institution{CREST - The Centre for Research on Engineering Software Technologies, The University of Adelaide}
\city{Adelaide}
\country{Australia}}
\affiliation{\institution{Cyber Security Cooperative Research Centre, Australia}
\city{}
\country{}}
\email{triet.h.le@adelaide.edu.au}

\author{Xiaoning Du}
\affiliation{\institution{Monash University}
\city{Melbourne, Victoria}
\country{Australia}}
\email{xiaoning.du@monash.edu}

\author{M. Ali Babar}
\affiliation{\institution{CREST - The Centre for Research on Engineering Software Technologies, The University of Adelaide}
\city{Adelaide}
\country{Australia}}
\affiliation{\institution{Cyber Security Cooperative Research Centre, Australia}
\city{}
\country{}}
\email{ali.babar@adelaide.edu.au}

\begin{abstract}

Collecting relevant and high-quality data is integral to the development of effective Software Vulnerability (SV) prediction models.
Most of the current SV datasets rely on SV-fixing commits to extract vulnerable functions and lines.
However, none of these datasets have considered latent SVs existing between the introduction and fix of the collected SVs.
There is also little known about the usefulness of these latent SVs for SV prediction.
To bridge these gaps, we conduct a large-scale study on the latent vulnerable functions in two commonly used SV datasets and their utilization for function-level and line-level SV predictions.
Leveraging the state-of-the-art SZZ algorithm, we identify more than 100k latent vulnerable functions in the studied datasets.
We find that these latent functions can increase the number of SVs by 4$\times$ on average and correct up to 5k mislabeled functions, yet they have a noise level of around 6\%.
Despite the noise, we show that the state-of-the-art SV prediction model can significantly benefit from such latent SVs.
The improvements are up to 24.5\% in the performance (F1-Score) of function-level SV predictions and up to 67\% in the effectiveness of localizing vulnerable lines.
Overall, our study presents the first promising step toward the use of latent SVs to improve the quality of SV datasets and enhance the performance of SV prediction tasks.

\end{abstract}



\maketitle

\section{Introduction}

Software Vulnerabilities (SVs) pose significant risks to the security and reliability of software systems.
With the increasing scale, complexity, and number of software applications, the effective detection of SVs has become an urgent need~\cite{hanif2021rise}.
Learning-based methods have garnered growing attention for automating the SV detection process~\cite{ghaffarian2017software,lin2020software}.
Recently, deep learning models have been increasingly used for the task, particularly the identification of potentially vulnerable functions (e.g.,~\cite{zhou2019devign,le2020deep,li2021vulnerability,fu2022linevul}).
The successful development of these SV prediction models heavily depends on the availability of high-quality SV datasets~\cite{croft2022data}.

There has been an increasing number of datasets of SVs in code functions~\cite{lin2022vulnerability}.
One of the most widely used approaches for curating such datasets is to leverage Vulnerability-Fixing Commits (VFCs) and label the code before the fixes as vulnerable~\cite{croft2023data}.
These VFCs are reported on security/SV advisories or can be obtained directly from project repositories on GitHub.
However, SVs may have been introduced long before they are fixed and exist across multiple software versions~\cite{meneely2013patch}.
Between the introduction and fix of a vulnerable function, many other commits may have modified the function without changing its vulnerable part, creating different vulnerable versions of the original function, so-called \textit{latent SVs}.
Unfortunately, current SV datasets have considered/included only vulnerable functions in fixing commits, not in prior commits, and thus, they have overlooked these latent SVs.

We posit that latent SVs, though being underutilized in the literature, can benefit the datasets and SV prediction.
In other words, they may be the valuable ``\textit{hidden gems}'' that can greatly impact the field of SV prediction.
Firstly, latent vulnerable functions may be mistakenly labeled as non-vulnerable due to a lack of awareness about latent SVs in existing datasets. 
The current practice involves collecting functions as non-vulnerable if they were not patched by VFCs (e.g.,~\cite{zhou2019devign,fan2020ac}).
Yet, these functions can be actually vulnerable if they contain the vulnerable parts of vulnerable functions fixed in respective VFCs.
Using latent SVs can address such mislabeling, which improves the quality of SV datasets as well as the training, optimization, and evaluation of SV prediction models~\cite{croft2023data}.
Secondly, incorporating latent SVs serves as a possible way to increase the overall quantity and the diversity of real-world code patterns of vulnerable functions in an SV dataset.
This can lead to a more comprehensive representation of SVs, which potentially enables SV prediction models to capture a wider range of security issues.

Despite the potential benefits, accurate identification of latent SVs in real-world projects is challenging.
Latent SVs require introducing commits (first vulnerable versions) of SVs, which have been commonly detected using the SZZ algorithm~\cite{borg2019szz}. 
For a vulnerable function, SZZ starts from the VFC and traces back the modifications, i.e., vulnerable code lines, made to the function until reaching the commit(s) adding the vulnerable lines.
Using the SZZ outputs, we can identify different historical versions between the introducing and fixing commits of the vulnerable function and collect them as latent vulnerable functions.
Formally, we define these latent SVs as ``\textit{vulnerable functions that are different previous versions of original vulnerable functions fixed in VFCs and contain semantically same vulnerable lines modified in the respective fixes}" (see \fig~\ref{fig:latent_sv}).
Our study also refers to these functions as \textit{SZZ-based latent SVs}.
Nevertheless, SZZ has been known to produce false positives~\cite{herbold2022problems}, which can introduce noise to SZZ-based latent SVs.
The aforementioned conflicting nature (data addition/correction capabilities vs. noise) raises an important question ``\textbf{\textit{To what extent do SZZ-based latent SVs affect SV prediction?}}''
To the best of our knowledge, there has been no study on such SZZ-based latent SVs in existing SV datasets and their use for SV prediction.

To bridge these gaps, we conduct a large-scale empirical study on the quality of SV datasets for SV prediction through the lens of SZZ-based latent vulnerable functions.
We customize the state-of-the-art \vszz{} algorithm~\cite{bao2022v} to identify and analyze 100,236 latent vulnerable functions from 24,871 vulnerable functions fixed in VFCs collected from the two popular datasets, namely \bigvul{}~\cite{fan2020ac} and \devign{}~\cite{zhou2019devign}.
Then, we evaluate their impacts on the state-of-the-art \linevul{} model~\cite{fu2022linevul} for function-level and line-level SV predictions.
Our key \textbf{contributions} can be summarized as follows:

\begin{itemize}
    \item We are the first to uncover and characterize latent vulnerable functions in the current SV datasets based on the latest SZZ algorithm.
    We show that 90\% of these latent SVs are missing in the datasets and they can correct up to 5k vulnerable functions mislabeled as non-vulnerable.
    Yet, around 6\% of these latent SVs are noisy.
    \item
    We demonstrate the positive impacts of using SZZ-based latent SVs in the studied datasets for SV prediction. We observe a rise of up to 24.5\% in F1-Score for function-level SV prediction and an effectiveness increase of up to 67\% for line-level SV prediction.
    The models with SZZ-based latent SVs can also identify up to 80\% of latent SVs that cannot be extracted based on SZZ outputs, around 2.5$\times$ better than the baseline models without latent SVs.
    \item We share our data and code for future research at~\cite{reproduction_package_msr2024_latent_sv}.
\end{itemize}

\noindent Our findings emphasize the value of including latent SVs to improve SV datasets and in turn downstream data-driven SV prediction.

\noindent \textbf{Paper structure}. Section~\mbox{\ref{sec:background}} introduces SV data labeling and the need for latent SVs. Section~\mbox{\ref{sec:rqs}} describes the three research questions. Section~\mbox{\ref{sec:data_collection}} presents the datasets and methods for collecting latent SVs to answer these questions. Sections~\mbox{\ref{sec:rq1_results}} and~\mbox{\ref{sec:rq2_rq3_results}} present the research methods and the results of each question. Section~\mbox{\ref{sec:discussion}} discusses the findings and the threats to validity. Section~\mbox{\ref{sec:conclusions}} concludes the study.

\begin{figure}[t]
    \centering
    \includegraphics[trim={19.5cm 4.3cm 27cm 4.3cm},clip,width=\columnwidth,keepaspectratio]{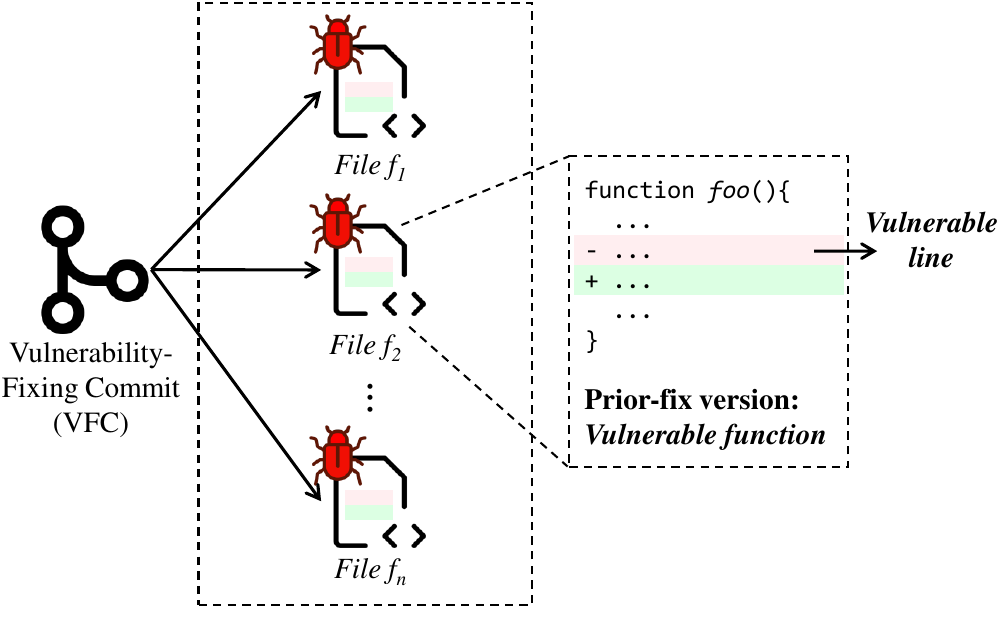}
    \caption{Procedure of current datasets for collecting vulnerable functions/lines from vulnerability-fixing commits.}
    \label{fig:data_collection}
\end{figure}

\section{Background and Motivation}\label{sec:background}

\subsection{SV Data Labeling in Current Datasets}
\label{subsec:data_labeling}

Vulnerable artifacts (i.e., functions/lines) in most of the existing SV datasets have been labeled from \textit{Vulnerability-Fixing Commits} (VFCs)~\cite{li2017large}, as illustrated in \fig~\ref{fig:data_collection}. Datasets following this data collection procedure curate \textit{real-world} SVs. This is particularly useful for the development of realistic downstream SV prediction models that can detect SVs in practical settings.
Specifically, VFCs are usually reported alongside the SVs in security advisories like National Vulnerability Database (NVD)~\cite{nvd_website}; in such cases, they are explicitly confirmed by developers/security experts. Some studies (e.g.,~\cite{chakraborty2021deep,zhou2019devign}) did not extract VFCs that had been reported/acknowledged by experts; rather, they automatically searched for VFCs. The automatic search has been performed by finding commit messages containing pre-defined SV-related keywords.
We consider the representative SV datasets utilizing both approaches of collecting VFCs, i.e., \bigvul{} using the NVD data and \devign{} relying on pre-defined SV keywords (see section~\ref{subsec:sv_datasets}).

Most datasets have extracted vulnerable functions and lines based on the common assumption that prior-fix code containing changes in VFCs is vulnerable~\cite{zhou2019devign,nguyen2019deep,li2021vulnerability,chakraborty2021deep,fu2022linevul} (see \fig~\ref{fig:data_collection}).
All the prior-fix files modified in a VFC are first considered to be vulnerable files. In each of the identified vulnerable files, prior-fix functions that contain the changed lines are then marked as \textit{vulnerable functions}. The modified lines themselves are labeled as \textit{vulnerable lines}. Many of the existing studies (e.g.,~\cite{zhou2019devign,fu2022linevul}) have extracted only deleted lines, \textit{not} added lines, as vulnerable. This decision has been made because of the challenge to automatically and accurately locate the vulnerable lines in the prior-fix functions leading to the added code.
We selected the function and line levels of granularity because they can help practitioners effectively and efficiently tackle SVs in practice~\cite{zhou2019devign,li2021vulnerability,fu2022linevul}.

\begin{figure}[t]
    \centering
    \includegraphics[width=0.99\columnwidth,keepaspectratio]{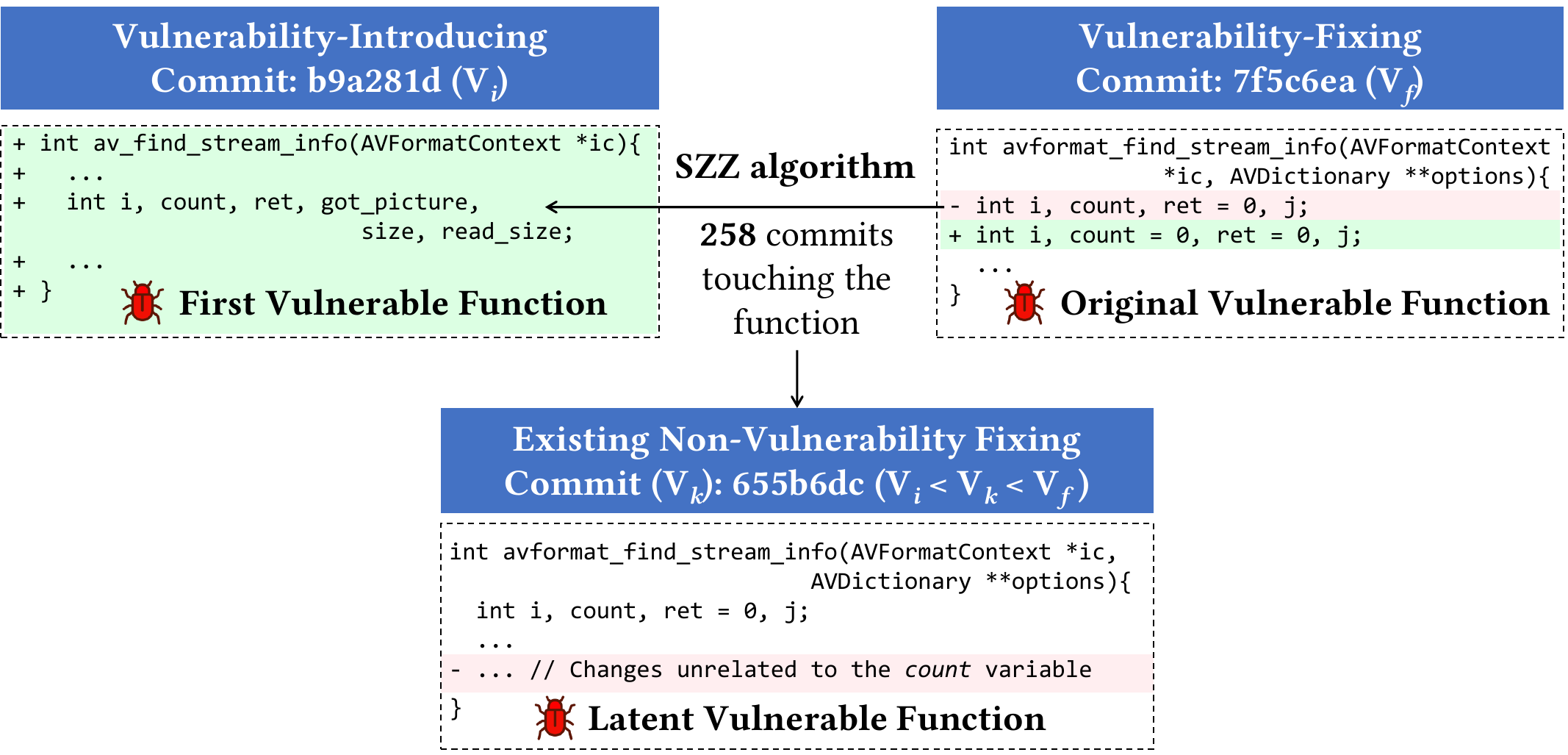}
    \caption{An exemplary latent vulnerable function existing between its introduction and fixing commits in the FFmpeg project. \textbf{Note}: This latent vulnerable function was originally labeled as non-vulnerable, as it belonged to a non-vulnerability fixing commit of the \devign{} dataset~\cite{zhou2019devign}. Its name was changed in one of the intermediate commits.}
    \label{fig:latent_sv}
\end{figure}

\subsection{Missing Consideration of Latent SVs}
\label{subsec:latent_svs_motivation}

There have been a growing number of datasets for vulnerable functions/lines~\cite{croft2023data}; however, to the best of our knowledge, none of them have considered \textit{latent SVs} when collecting the data.
Latent SVs occur because SVs are usually not fixed soon after they are introduced in code commits as part of continuous integration~\cite{arani2023sok}. Previous studies (e.g.,~\cite{le2021deepcva,meneely2013patch}) showed that SVs had been actually introduced a long time, more than 1,000 days on average, before they were fixed in VFCs, posing significant yet hidden security risks to the affected projects.
\fig~\ref{fig:latent_sv} gives an example of a latent vulnerable function. Here, the function \code{avformat\_find\_stream\_info} is vulnerable in the FFmpeg project of the \devign{} dataset~\cite{zhou2019devign}. This function contains a vulnerable line with the uninitialized variable \code{count}, which was modified/fixed in the VFC \textit{7f5c6ea} by adding the zero-value initialization.
The SV (vulnerable line) was introduced in the commit \textit{b9a281d}, which was 5,121 days before the fix.
Any different versions of this function between its introducing and fixing commits are considered latent vulnerable functions because the vulnerable line still exists in these versions.
One of the intermediate commits touching this function, i.e., \textit{655b6dc}, contains one of such latent vulnerable functions.
Latent SVs can change the characteristics (e.g., class distributions) of existing SV datasets and in turn affect the performance of downstream SV prediction models.

Firstly, latent SVs can increase SV data size, i.e., the number of SVs.
In real-world projects, between the SV introducing and fixing commits of a vulnerable function, many other changes/commits have been made to the function to fix bugs and/or add new features.
For instance, using the SZZ algorithm~\cite{bao2022v}, we found that 258 different commits modified the function \code{avformat\_find\_stream\_info}, as illustrated in \fig~\ref{fig:latent_sv}, yet kept the vulnerable line unchanged between its introduction and fix.
Accordingly, there were 258 unique latent vulnerable functions corresponding to the vulnerable function.
If these latent vulnerable functions are included in an SV dataset, the proportion of the vulnerable class will be boosted significantly. Such an increase can help prediction models learn more diverse real-world SV patterns and potentially improves the resulting predictive performance~\cite{croft2022data}.

Secondly, latent SVs can correct the labels of functions originally labeled as non-vulnerable. These functions were non-vulnerable initially, mostly because they were not included/modified in VFCs.
For example, the latent vulnerable function in \fig~\ref{fig:latent_sv} was originally labeled as non-vulnerable in the \devign{} dataset~\cite{zhou2019devign}. According to the data collection heuristics, this function was changed in a non-SV fixing commit \textit{655b6dc} and thus was not vulnerable. Still, the uninitialized variable from the original function indicates the SV in this prior function. Lacking the additional information of latent SVs, it would be non-trivial to determine that this function is vulnerable without human intervention~\cite{croft2022noisy}.
Unfortunately, there is no perfect solution to label latent SVs at scale~\cite{croft2022noisy,bao2022v}. Therefore, the extent of the aforementioned benefits largely depends on how accurately latent SVs are identified.

The quality of latent SVs depends on the label accuracy of SV-introducing commits.
Manual identification of these latent SVs is ideal, but it is impractical to manually label whole datasets~\cite{croft2022noisy}. Also, affected versions of SVs reported on security advisories are usually inaccurate and/or incomplete~\cite{anwar2021cleaning}. Instead, the SZZ algorithm~\cite{sliwerski2005changes} and its variants (e.g., V-SZZ~\cite{bao2022v} we used in \fig~\ref{fig:latent_sv}) have been considered the standard method to automatically trace SV-introducing commits from VFCs~\cite{bao2022v}. Yet, they still produce false-positive SV-introducing commits~\cite{fan2021impact}, which can lead to mislabeled/noisy latent SVs.\footnote{Noisy latent SVs (actually non-vulnerable functions) are found/discussed in section~\ref{sec:rq1_results}, confirming that SZZ-generated false positives result in incorrect latent SVs.}
Many studies have investigated the quality and improvement of SV-introducing commits output by SZZ (e.g.,~\cite{kim2006automatic,da2016framework,neto2018impact}) as well as using such commits directly for empirical analysis (e.g.,~\cite{meneely2013patch,bosu2014identifying,paul2021security}) and/or just-in-time/commit-level SV prediction~\cite{perl2015vccfinder,yang2017vuldigger,lomio2022just}.
There has been no investigation of the quality and prevalence of \textit{SZZ-based latent SVs} in existing SV datasets. Moreover, given their aforementioned large-scale yet noisy nature, little is known about how much latent SVs can benefit downstream SV prediction models.
To the best of our knowledge, we are the first to empirically study the quality and impact of latent vulnerable functions on function-level and line-level SV predictions.

\section{Research Questions}
\label{sec:rqs}

We answer three Research Questions (RQs) to explore SZZ-based SVs in SV datasets and their impacts on function-level and line-level SV predictions.
Note that we do not study latent non-vulnerable functions as it is extremely hard to guarantee the absence of SVs in source code~\cite{weinberg2008perfect}.

\textbf{RQ1: What are the quality and prevalence of SZZ-based latent SVs in existing SV datasets?}
The SZZ outputs can uncover latent vulnerable functions, but little is known about the nature of such SZZ-based latent SVs in existing SV datasets.
Prior studies have not incorporated such latent SVs into the SV datasets being used for training and evaluating state-of-the-art SV prediction models like \linevul{}~\cite{fu2022linevul}. RQ1 investigates the extent to which latent SVs are accurately labeled based on the outputs of the state-of-the-art \vszz{} algorithm~\cite{bao2022v} and their overlaps with SVs already curated in existing SV datasets.
RQ1 aims to distill the quality and prevalence of these latent SVs, which can provide a better understanding of how latent SVs can be used for enhancing SV prediction models.

\textbf{RQ2: How do SZZ-based latent SVs impact function-level SV prediction?}
SZZ-based latent SVs have values yet are noisy.
They provide additional data to the vulnerable class that can boost SV predictive performance.
RQ2 explores the impact of these SVs on \linevul{}~\cite{fu2022linevul}, the state-of-the-art model for function-level SV prediction.
Given the presence of noise in latent SVs, we also consider noise-aware models powered by noise-reduction techniques and compare them with noise-unaware models for the task.
RQ2 findings can shed light on the level of necessity/importance of considering latent vulnerable functions when curating SV datasets for improving the performance of function-level SV prediction. 

\textbf{RQ3: How do SZZ-based latent SVs impact line-level SV prediction?}
Line-level SV prediction is increasingly important as it can reveal model interpretability and cost-to-inspect, which are critical requirements for real-world usage of SV prediction models~\cite{jiarpakdee2021practitioners,jiarpakdee2020empirical}. RQ3 studies whether latent SVs can enhance line-level SV prediction using the \linevul{}~\cite{fu2022linevul} model.
Like RQ2, we also compare noise-aware models with noise-unaware ones for the task.
RQ3 findings can unveil the usefulness of latent SVs for improving the effectiveness of localizing vulnerable lines, which enables better industrial adoption of current SV prediction models.

\section{Data Collection}
\label{sec:data_collection}

This section describes the data collection process we used for investigating the quality and prevalence of latent SVs (vulnerable functions) and their use for SV prediction. Firstly, we obtained vulnerable and non-vulnerable functions based on the code commits in the existing SV datasets (see section~\ref{subsec:sv_datasets}). From the vulnerable functions, we extracted latent vulnerable functions using the outputs of the state-of-the-art V-SZZ algorithm~\cite{bao2022v} (see section~\ref{subsec:latent_svs_szz}).

\subsection{Studied SV Datasets}
\label{subsec:sv_datasets}

The data we curated for answering the three RQs were built on top of the \bigvul{}~\cite{fan2020ac} and \devign{}~\cite{zhou2019devign} datasets. Both datasets relied on information from Vulnerability-Fixing Commits (VFCs) hosted on GitHub to determine whether a function was vulnerable or not. The details of the datasets used in this study are given below.

\textbf{\bigvul{}}~\cite{fan2020ac} gathered vulnerable and non-vulnerable functions from VFCs provided in the CVE Details database~\cite{cve_details}. Any functions containing lines changed in the collected VFCs were labeled as vulnerable. Then, all remaining functions that were not modified in the files touched by VFCs were labeled as non-vulnerable.

\textbf{\devign{}}~\cite{zhou2019devign} collected vulnerable functions from VFCs, but unlike \bigvul{}, it obtained non-vulnerable functions from non-VFCs, i.e., commits not fixing SVs. The \devign{} authors used a keyword-matching approach to label whether a commit was a VFC or not. They also manually validated the VFCs to ensure data accuracy. Then, functions changed in the VFCs were labeled as vulnerable, and functions changed in the non-VFCs were labeled as non-vulnerable.

We chose these two datasets because of the three reasons:

\begin{enumerate}[leftmargin=*]
  
    \item \bigvul{} and \devign{} are among the most popular SV datasets, together with ReVeal (FFmpeg+Qemu)~\cite{chakraborty2021deep}, each with more than 100 citations. However, ReVeal does not contain explicit links to the VFCs for extracting latent SVs~\cite{reveal_issue}, making this dataset unsuitable for our study.

    \item These datasets contain real-world and diverse SVs, which can help increase the relevance of our investigations. \bigvul{} extracted data from 300+ open-source C/C++ projects. \devign{} consisted of FFmpeg and OpenSSL, which are well-maintained projects with real-world impacts and usages.

    \item \bigvul{} and \devign{} provide different characteristics and distributions of SVs. \bigvul{} and \devign{} differ in the methods to collect the non-vulnerable functions, as mentioned above.
    Accordingly, \bigvul{} has an imbalanced ratio between vulnerable and non-vulnerable functions, while \devign{} possesses a nearly balanced distribution between the two classes. Such differences help enhance the generalizability of our study.

\end{enumerate}

\begin{figure*}[t]
    \centering
    \includegraphics[width=0.99\textwidth,keepaspectratio]{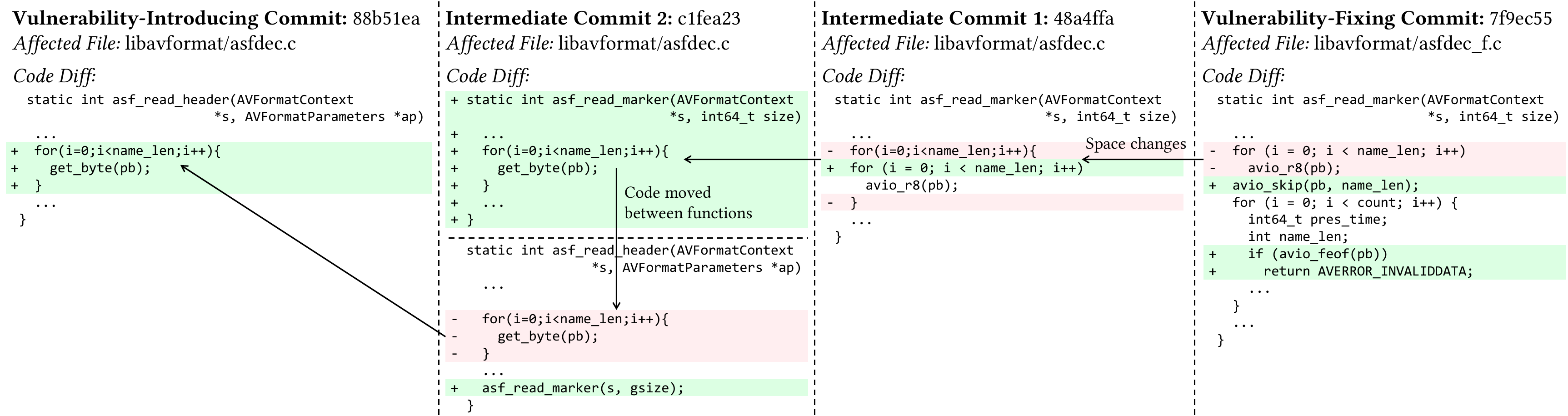}
    \caption{An example of using \vszz{} to identify the Vulnerability-Introducing Commit from a Vulnerability-Fixing Commit with intermediate commits performing only refactoring. \textbf{Notes}: The arrows show the tracking of the original vulnerable line \code{for (i = 0; i < name\_len; i++)} across commits. The commits were taken from the FFmpeg project of the \devign{} dataset.}
    \label{fig:vszz_example}
\end{figure*}

\noindent \textbf{Adaptation of the datasets to our study}.
Our data curation required explicit links to VFCs, so we would discard a commit in the original datasets if it were no longer available/accessible.
It is also important to note that we excluded the Chromium project~\cite{chromium} from \bigvul{} as VFCs extracted from this project were shown to be mostly inaccurate~\cite{croft2023data}, which would affect the reliability of our findings.
The numbers of the collected vulnerable and non-vulnerable functions are given in \tab~\ref{tab:data_statistics}.

\begin{table}[t]
\fontsize{8.5}{9.5}\selectfont

  \centering
  \caption{The numbers of vulnerable functions, non-vulnerable functions, and SZZ-based latent vulnerable functions in the \bigvul{} and \devign{} datasets used in this study.}
 \begin{tabular}{lccc|c}
    \hline
    \multirowcell{3}[0ex][l]{\textbf{Dataset}} & \multicolumn{4}{c}{\textbf{Data composition}}\\
    \cline{2-5}
    & \makecell{\textbf{Vuln.} \\ \textbf{functions}} & \makecell{\textbf{Non-vuln.}\\ \textbf{functions}} &  \makecell{\textbf{\% Vuln.}\\ \textbf{functions}} & \makecell{\textbf{Latent vuln.}\\ \textbf{functions}} \\
    \hline
    \textbf{\bigvul{}} & 5,554 & 151,801 & 3.5\% & 34,700 \\
    \textbf{\devign{}} & 19,317 & 24,369 & 44.2\% & 65,536 \\
    \hline
    \textbf{Total} & 24,871 & 176,170 & 12.4\% & 100,236 \\
    \hline
    \end{tabular}
  \label{tab:data_statistics}
\end{table}

\subsection{SZZ-Based Identification of Latent SVs}
\label{subsec:latent_svs_szz}

For each of the original vulnerable functions\footnote{In this paper, we also refer to vulnerable functions in the two datasets as \textit{original vulnerable functions} from which \textit{latent vulnerable functions} can be curated.} in section~\ref{subsec:sv_datasets}, we used \vszz{}~\cite{bao2022v}, the state-of-the-art SZZ algorithm, to identify latent vulnerable functions.
Given a vulnerable function with changed lines, V-SZZ~\cite{bao2022v} outputs the commits introducing/adding each of the vulnerable lines in the input function, namely Vulnerability-Introducing Commits (VICs).
Fundamentally, an SZZ algorithm~\cite{sliwerski2005changes} selects last commits that made functional modifications to the source code lines deleted or modified to address an SV in a VFC (see \fig~\ref{fig:latent_sv}), using tools like \code{git blame}.
Improving upon prior SZZ variants, \vszz{} utilizes a novel line mapping algorithm to trace VICs more accurately. This line mapping algorithm performs code similarity checking, e.g., more than 75\% similar based on edit distance, to map after-commit and prior-commit versions of vulnerable code line(s). This is particularly useful when the syntax of vulnerable code changes, but its semantics (vulnerable parts) remain unchanged; e.g., a vulnerable code line underwent cosmetic changes or was moved between the files and/or functions. Because of this enhanced capability, we chose \vszz{} to increase the accuracy of uncovering VICs and in turn latent vulnerable functions.

The example in \fig~\ref{fig:vszz_example} illustrates how \vszz{} uses the line mapping algorithm to extract the VIC from the VFC \textit{7f9ec55} in the FFmpeg project of the \devign{} dataset. The goal here is to identify the VIC of the vulnerable line \code{for (i = 0; i < name\_len; i++)}.
For this case, \vszz{} correctly labeled the commit \textit{88b51ea} as the VIC.
Using the line mapping algorithm, \vszz{} could track the line in the project history even when the name of the enclosing file changed from \code{libavformat/asfdec\_f.c} to \code{libavformat/asfdec.c}. However, this case would have been missed by some SZZ variants that do not consider file name change when tracing VICs like B-SZZ~\cite{sliwerski2005changes}.
Moreover, \vszz{} also managed to skip the two incorrect VICs of this line, namely \textit{48a4ffa} and \textit{c1fea23}, which had also modified the vulnerable line.
The former commit \textit{48a4ffa} was the last one modifying the vulnerable line before the VFC, but it did not perform any functional change (i.e., only space changes).
The latter commit \textit{c1fea23} extracted the code, including the vulnerable line in the function \code{asf\_read\_header}, and moved it to the new function \code{asf\_read\_marker} to enhance code maintainability. This \textit{ExtractMethod} refactoring operation means that the vulnerable line was moved between the two functions~\cite{mens2004survey}.
Without the line mapping used in \vszz{}, prior SZZ variants might assign a false-positive VIC to this case by selecting either of these two refactoring-only commits.
The output (VIC) of \vszz{} enabled us to determine latent vulnerable versions (i.e., the functions existing between the first vulnerable and fixed ones) of each original vulnerable function.

\noindent \textbf{Adaptation of \vszz{} to identify latent vulnerable functions}.
\vszz{}~\cite{bao2022v} does not directly identify latent vulnerable functions, so we customized its output for our study.
For each original vulnerable function in a VFC, we identified all vulnerable lines in the function and then applied \vszz{} to find the commits that had introduced these lines, i.e., VICs.
Among the extracted VICs, we took the earliest one based on commit date as the version when the original vulnerable function started to become vulnerable.
From the commit(s) between the VFC and VIC that touched/modified the original function, we obtained all of the versions of the function prior to each commit.
Accordingly, these versions were labeled as \textit{latent vulnerable functions} because they still contained the same or semantically similar vulnerable lines as the original vulnerable function.
When labeling the latent vulnerable functions, we considered their function name change(s) to maximize completeness.
Note that a few of these latent vulnerable functions may be actually non-vulnerable (e.g., due to the imprecisions of \vszz{}), which are discussed in section~\ref{sec:rq1_results}.
After the aforementioned process, we identified 100k+ latent vulnerable functions, i.e., \textit{34,700} and \textit{65,536} ones from the \bigvul{} and \devign{} datasets, respectively, as shown in \tab~\ref{tab:data_statistics}.
The smaller number of latent SVs in \bigvul{} is likely because this dataset had fewer original SVs due to missing many VFCs~\cite{sawadogo2022sspcatcher,nguyen2022hermes}.
In contrast, \devign{} extensively searched for and manually validated VFCs, resulting in much more original SVs.
For instance, the same FFmpeg and Qemu projects in \bigvul{} only had 84 and 1 VFCs, respectively, much smaller than those (5,962 and 4,932) in \devign{}.

\section{Quality and Prevalence of Latent SVs in Existing Datasets (RQ1)}
\label{sec:rq1_results}

This section describes the investigation of characteristics of latent vulnerable functions collected based on \vszz{} with respect to original vulnerable and non-vulnerable functions.
Firstly, we manually inspected the quality of these latent SVs in section~\ref{subsec:quality_latent_svs}.
Based on the manually identified noise level ($\approx$6\%), we then distill the prevalence of these latent SVs in existing SV datasets in section~\ref{subsec:prevalence_latent_svs}.

\subsection{Quality of Latent SVs}
\label{subsec:quality_latent_svs}

\subsubsection{\textbf{Methods of Manual Validation}}

We manually validated the labels of the vulnerable functions identified based on the outputs of V-SZZ because this SZZ algorithm is imperfect~\cite{bao2022v}.
It should be noted that we did not validate whether each \vszz{} output was a genuine SV-introducing commit as already done in~\cite{bao2022v}, rather we checked whether it revealed a correct latent SV.
Our validation was different from~\cite{bao2022v} as an incorrect SV-introducing commit (e.g., not the earliest one of an SV) could still generate correct latent SVs.
Our results can provide insights into the quality of latent vulnerable functions that have been missed by the existing datasets and their potential usage for SV prediction.
We could not manually validate all the identified latent vulnerable functions as it would be too time-consuming. Instead, we decided to work with a subset that could be completed within a reasonable amount of time and effort, which is also a common practice in the Software Engineering research~\cite{baltes2022sampling}.
Two of the authors, with at least five years of experience in the Software Engineering and SV areas, independently conducted manual validation of the labels of 70 latent vulnerable functions collected from each of the \bigvul{} and \devign{} datasets.
Note that the sample size (70) was of statistical significance with 90\% confidence and 10\% margin error~\cite{cochran2007sampling}.
We did not sample any VFC more than once to ensure that the results were not biased toward that commit and thus more representative of the commits in each dataset.

We followed a two-step process to validate the label of each latent vulnerable function.
We first read the original vulnerable function, its line(s) fixed, and the commit message in the respective VFC to comprehend the line(s) related to the SV, i.e., original vulnerable lines. 
We then read and comprehended the code in the latent vulnerable function, and the introducing commit labeled by \vszz{}, to determine whether the latent vulnerable function had any of the original vulnerable lines.
We marked a latent function as vulnerable (true positive) only if we deemed it contained the same SV fixed in the VFC by checking the vulnerable lines and the contextual code in the current and relevant functions.
Latent functions without the same SV as in the original vulnerable function, but were subjected to a different SV, are out of the scope of our study.

We take the function \code{avformat\_find\_stream\_info} in \fig~\ref{fig:latent_sv} to illustrate the labeling process. Here, we found that the function in the intermediate commit \textit{655b6dc} was an earlier version of the original function fixed in the VFC \textit{7f5c6ea} (i.e., having the same name), and also contained the same vulnerable line (i.e., the \code{count} variable without initialization). We also confirmed that the changes in the intermediate commit did not affect the SV semantics. Leveraging such information, we would label this case as a true-positive latent vulnerable function. Following the independent labeling, the two labelers discussed any disagreements to reach consensus.
The Cohen's Kappa
($\kappa$) inter-rater reliability score~\cite{mchugh2012interrater} between the two labelers was 0.93, i.e., indicating ``almost perfect'' agreement~\cite{hata20199}.

We also documented the reasoning behind our labels and then performed thematic analysis~\cite{braun2006using} of the labeling reasoning to identify the patterns/causes of the false positives of SZZ-based latent vulnerable functions.
Such patterns would give insights into potential ways of leveraging these latent functions for SV prediction in section~\ref{sec:rq2_rq3_results}.
The manual labeling and analysis of latent vulnerable functions were considerably labor-intensive, taking approximately 90 man-hours.
More details can be found at~\cite{reproduction_package_msr2024_latent_sv}.

\subsubsection{\textbf{Results of Manual Validation}}

Among the 140 latent vulnerable functions we manually validated, only nine (6.4\%) of them were false positive/non-vulnerable, including three and six false positives for \bigvul{} and \devign{}, respectively.
Based on our thematic analysis of these false positives, we discovered two key patterns: \textit{incorrect line mapping} and \textit{changed code context} between the original and latent vulnerable functions.

\noindent \underline{\textit{Incorrect line mapping}}. This false-positive pattern directly resulted from the line mapping algorithm used by \vszz{}~\cite{bao2022v}, as described in  section~\ref{subsec:latent_svs_szz}.
This mapping algorithm assumes that similarly-looking code shares similar functionality and the same SV. From our manual labeling and analysis, this assumption is reasonable for many cases like the one in \fig~\ref{fig:vszz_example}, but it is not always accurate, leading to false-positive latent vulnerable functions.
For example, the function \code{mov\_write\_header} had a vulnerable line containing \code{mov->mode == MODE\_ISM} removed/fixed in the VFC \textit{a9553bbb} of the FFmpeg project of the \devign{} dataset~\cite{zhou2019devign}. The line implies that the SV was related to the video format \textit{ISM} used by FFmpeg~\cite{ffmpeg_video_format}. Based on the \vszz{} output of this original vulnerable function, a latent vulnerable function was extracted from an intermediate commit \textit{0c716ab}. Actually, this latent function was not vulnerable, and thus false-positive, because it did not contain any code related to \textit{ISM}. It only had code related to other video formats of FFmpeg such as \textit{3GP}, \textit{MOV}, \textit{MP4}, and \textit{PSP}. With a further investigation on this function, we discovered that \vszz{} correctly detected the introducing commit \textit{187023f6} for the deleted line containing \code{mov->mode == MODE\_ISM}, but then incorrectly mapped this vulnerable line to a prior-commit line containing \code{mov->mode == MODE\_PSP} in the same commit. These lines were nearly identical with only two-character differences (e.g., \textit{\underline{I}S\underline{M}} vs. \textit{\underline{P}S\underline{P}}), but they actually belonged to two completely different video formats.

\noindent \underline{\textit{Changed code context}}.
Any false-positive latent vulnerable functions of this type generally do not have the same code context causing the SVs as in the original vulnerable functions. These cases mainly involve SVs that do not only stem from the deleted line(s) in the respective VFCs but also surrounding code in the affected functions. For instance, the function \code{bdrv\_close} contained a vulnerable line \code{bs->drv->bdrv\_close(bs)} fixed in the VFC \textit{9a7dedb} in the Qemu project of the \devign{} dataset~\cite{zhou2019devign}. This function call could lead to a potential double-free SV because it might revoke the child nodes of the block driver (\code{bs}), though its child nodes were already removed in the code block preceding it.
However, the function labeled as latent vulnerable from the commit \textit{19cb373} based on the \vszz{} output did not contain the code block. Thus, this latent function was actually non-vulnerable and false-positive.

Overall, latent vulnerable functions have decent quality with around 6\% noise. The distilled noise usually involves project-specific code semantics and requires a deep code understanding to remove.

\subsection{Prevalence of Latent SVs}
\label{subsec:prevalence_latent_svs}

\subsubsection{\textbf{Methods}}

For each of the \bigvul{} and \devign{} datasets, we analyzed overlapping (with exact same code) and divergent cases (with different code) between all the latent vulnerable functions and all the original vulnerable and non-vulnerable functions.
The overlapping functions could be originally labeled as vulnerable or non-vulnerable.
The functions that were originally labeled as vulnerable mean that the existing SV datasets did not miss these cases.
The overlapping functions that were originally labeled as non-vulnerable imply that these cases were potentially mislabeled by the data collection methods used in the SV datasets.
The divergent cases are vulnerable functions that have been missed by the datasets.
Both scenarios where latent SVs add new functions and correct mislabeled functions can be leveraged to enhance SV prediction.

\begin{table}[t]
\fontsize{8.5}{9.5}\selectfont

  \centering
  \caption{The numbers of SZZ-based
  latent vulnerable functions that were originally labeled as vulnerable, originally labeled as non-vulnerable, or missing in the \bigvul{} and \devign{} datasets. \textbf{Note}: The percentages of row-wise values computed with respect to the total number of latent vulnerable functions in each dataset are in parentheses.}
 \begin{tabular}{lccc|c}
    \hline
    \textbf{Dataset} & \makecell{\textbf{Orginally} \\ \textbf{vulnerable}} & \makecell{\textbf{Originally}\\ \textbf{non-vuln.}} &  \makecell{\textbf{Missing}} & \textbf{Total}\\
    \hline
    \textbf{\bigvul{}} & 560 (1.6\%) & 519 (1.5\%) & 33,621 (96.9\%) & 34,700 \\
    \textbf{\devign{}} & 5,610 (8.6\%) & 4,551 (6.9\%) & 55,375 (84.5\%) & 65,536 \\
    \hline
    \textbf{Total} & 6,170 (6.2\%) & 5,070 (5.1\%) & 88,996 (88.7\%) & 100,236 \\
    \hline
    \end{tabular}
  \label{tab:latent_data_analysis}
\end{table}

\subsubsection{\textbf{Results}}

Table~\ref{tab:latent_data_analysis} shows that the majority ($\approx$90\%) of the identified latent vulnerable functions were missing/new in the \bigvul{} and \devign{} datasets.
With the demonstrated quality (i.e., noise level of $\approx$6\%) in section~\ref{subsec:quality_latent_svs}, many of these new latent functions were actually true positives/vulnerable and could significantly increase the numbers of SVs in the studied datasets by 4$\times$ on average. The increase in SV data size in turn raised the ratio of vulnerable functions to non-vulnerable ones.
After incorporating all the identified latent vulnerable functions and excluding the overlapping ones, the SV ratio in \bigvul{} could rise nearly six times from 3.5\% to 20.7\%, while the ratio in \devign{} would almost 1.8$\times$ from 44.2\% to 76.5\%.\footnote{With the low noise level, the exact values would be close to the presented ones.}

Around 11\% of the identified latent vulnerable functions overlapped with the original vulnerable and non-vulnerable functions in the studied datasets (see Table~\ref{tab:latent_data_analysis}).
There were 6.2\% (1.6\% for \bigvul{} and 8.6\% for \devign{}) of the latent SVs already included in the datasets, i.e., originally labeled as vulnerable.
This implies that the overlapping latent SVs were modified in some VFCs of \bigvul{} and \devign{}, but these VFCs might patch different SVs or partially fix the same SVs.
This also aligns with the existing finding that 22--33\% of bugs need more than one fix attempt~\cite{park2012empirical}.
Moreover, 5.1\% of these latent functions were originally labeled as non-vulnerable in the datasets.
The results indicate that the common assumptions, i.e., unchanged functions in VFCs (\bigvul{}) or changed functions in non-VFCs (\devign{}), being used for collecting non-vulnerable functions are not always accurate.
Unlike \bigvul{} and \devign{}, if a dataset included all functions, the ($\approx$89k) latent vulnerable functions normally considered as non-vulnerable (i.e., they were not part of known SV-fixing commits) would be mislabeled.
These values confirm that latent vulnerable functions not only add missing vulnerable functions to SV datasets but also correct many functions mislabeled as non-vulnerable.
Accordingly, we highlight a promising way to validate the quality of an SV dataset (e.g., finding missing and mislabeled functions) through the lens of latent SVs.

\begin{figure*}[t]
    \centering
    \includegraphics[trim={16cm 4cm 19cm 6cm},clip,width=\textwidth,keepaspectratio]{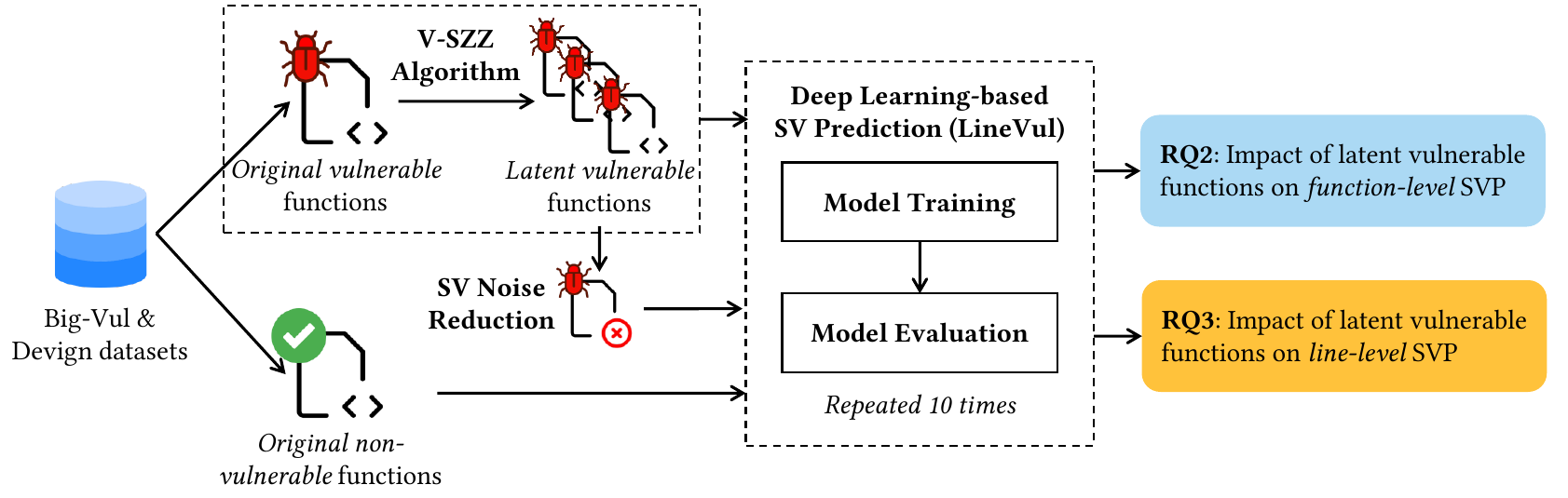}
    \caption{Workflow of evaluating the impacts of latent vulnerable functions on function-level and line-level SV predictions. \textbf{Note}: SVP stands for SV prediction.}
    \label{fig:model_building_method}
\end{figure*}

The aforementioned findings show that latent SVs substantially changed the data distributions of both classes, particularly the vulnerable one, of the studied SV datasets.
This also suggests that the incorporation of latent SVs can potentially impact the performance of function-level and line-level SV prediction, which is investigated in-depth in section~\ref{sec:rq2_rq3_results}.
We also assert that it is very challenging to detect and tackle accurately all the complex noise of latent SVs using manually defined rules.
Therefore, in section~\ref{sec:rq2_rq3_results}, we also explore automatic techniques for reducing noise in latent SVs and compare the performance between noise-aware and noise-unware SV prediction models using latent SVs.

\begin{tcolorbox}
\textbf{RQ1 Summary}.
SZZ-based latent SVs have a noise level of 6.4\%, mainly due to incorrect line mapping or changed code context. The noise involves complex code semantics and is challenging to remove.
SZZ-based latent SVs are also 90\% new and 4$\times$ the SV data size in existing datasets (\bigvul{} and \devign{}) and correct up to 5k mislabeled functions.

\end{tcolorbox}

\section{Impact of Latent SVs on SV Prediction Models (RQ2 \& RQ3)}
\label{sec:rq2_rq3_results}

In this section, we evaluate different ways and respective impacts of using latent vulnerable functions extracted from existing datasets for function-level (RQ2) and line-level (RQ3) SV prediction.

\subsection{Research Methods}

The methods used to answer RQs 2 and 3 are illustrated in \fig~\ref{fig:model_building_method}.
The first step is the preparation of three types of data, including original vulnerable and non-vulnerable functions as well as latent vulnerable functions for function-level and line-level SV predictions.
The original vulnerable and non-vulnerable functions were obtained from the \bigvul{} and \devign{} datasets, as described in section~\ref{subsec:sv_datasets}. The latent vulnerable functions were extracted from the original vulnerable functions based on the \vszz{} outputs, as described in section~\ref{subsec:latent_svs_szz}.
We removed duplicates in the original vulnerable, latent vulnerable, and non-vulnerable functions that only differed due to cosmetic changes (e.g., changes in spaces, tabs, and newlines). This removal was based on the recommendation that duplicate data would be likely to have a negative impact on SV prediction~\cite{croft2023data}.
To provide labels for line-level SV prediction, we leveraged the modified/deleted lines of these functions in respective VFCs as vulnerable lines, following the common practice recommended in the literature (e.g.,~\cite{jiarpakdee2020empirical,li2021vulnerability,fu2022linevul}).

The functions were input into \linevul{}~\cite{fu2022linevul}, the state-of-the-art model for both function-level (RQ2) and line-level (RQ3) SV prediction~\cite{steenhoek2023empirical}. The details of the \linevul{} model and its configurations used in our study are given in section~\ref{subsubsec:svp_algorithms}. Because RQ1 shows that latent vulnerable functions are noisy (see section~\ref{sec:rq1_results}), we also investigated various noise-reduction techniques in section~\ref{subsubsec:data_cleaning} with the aim of automatically tackling such noise.
We named the SV prediction models using noise-reduction as \textit{noise-aware models}, while the ones not using as \textit{noise-unaware models}.
We compared noise-aware models with noise-unaware models that incorporated latent vulnerable functions, as well as with baseline models that did not use latent SVs.
The technique (e.g., data splitting) and measures used for evaluating these models in RQ2/RQ3 are covered in sections~\ref{subsubsec:evaluation_technique} and~\ref{subsubsec:evaluation_measures}, respectively.

\subsubsection{\textbf{SV Prediction Model}}
\label{subsubsec:svp_algorithms}

We chose \linevul{}, a recent Deep Learning-based SV prediction model~\cite{fu2022linevul}, for its demonstrated performance superiority over previous baselines for both function-level and line-level prediction~\cite{steenhoek2023empirical}. \linevul{} relies on CodeBERT~\cite{feng2020codebert}, a high-performing pre-trained code embedding model based on the RoBERTa architecture~\cite{liu2019roberta}, to extract code feature representations that capture syntactic and semantic information.
In the case of \linevul{}, function-level predictions are generated using a Transformer-based architecture, followed by the identification of the most vulnerable lines within these functions using the attention scores obtained from the trained Transformer model.
The hyperparameters of \linevul{} we used were as follows: \textit{epochs}: 10, \textit{learning rate}: 1e-5, and \textit{feature embedding size}: 768. These hyperparameters were taken from the prior studies (e.g.,~\cite{steenhoek2023empirical,croft2023data}) leveraging \linevul{}  for SV prediction (without latent SVs).
We investigated values other than the selected ones, but we did not observe any significant improvement in performance.

\subsubsection{\textbf{Noise Reduction in SZZ-Based Latent Vulnerable Functions}}
\label{subsubsec:data_cleaning}

We investigated three noise-reduction techniques, i.e., (\textit{i}) \textit{Latest Introducing Commit}, (\textit{ii}) \textit{Self-Training}, and (\textit{iii}) \textit{Centroid-based Removal}, to remove incorrect SZZ-based latent vulnerable functions \textit{in the training sets}.
The noise removal potentially enhances SV predictive performance using latent vulnerable functions.

\noindent \textbf{Latest Introducing Commit (LIC)}.
This technique is \textit{data-based} and inspired by the first noise pattern of latent vulnerable functions in RQ1, i.e., incorrect line mapping in \vszz{}.
\vszz{} applies the line mapping algorithm only after the latest SV-introducing commit is found for a given vulnerable function (see \fig~\ref{fig:vszz_example}). Thus, LIC used the version(s) of the function after the latest introducing commit, based on commit date, which would not be affected by such noise.
In the illustrated case of incorrect line mapping in RQ1, any versions of the function \code{mov\_write\_header} following the commit \textit{187023f6} (the latest SV-introducing commit) until the VFC \textit{a9553bbb} still had \code{MODE\_ISM}, confirming the existence of the SV.
In addition, latent functions after the latest introducing commit would probably be more similar to the original vulnerable function than those prior to the commit because fewer changes/commits (i.e., closer to the VFC) were made to the function. This may help reduce the probability of the second noise pattern, i.e., changed code context.

\noindent \textbf{Self-Training (ST)}. ST is one of the most popular \textit{model-based} semi-supervised techniques that can assign labels to unlabeled data using the trained models on labeled data~\cite{chapelle2009semi}. Specifically, if a latent vulnerable function had a likelihood of more than 0.5 of being non-vulnerable, predicted by models learned from original data, we would mark this latent function as noisy and remove it.

\noindent \textbf{Centroid-based Removal (CR)}. CR is a commonly used \textit{feature-based} approach for detecting noisy samples in the Software Engineering domain (e.g.,~\cite{le2020puminer,tian2012identifying}). CR removes latent vulnerable functions having the feature vectors closer (with smaller cosine distances) to the \textit{centroid} (average feature vector) of the non-vulnerable class than that of the non-vulnerable class.

Noise-aware models based on the noise-reduction techniques used the same \linevul{} model. We did not apply noise reduction to the original data because that is beyond the scope of our study.

\subsubsection{\textbf{Evaluation technique}}
\label{subsubsec:evaluation_technique}
We employed a 10-round evaluation. In each round, we split the original vulnerable and non-vulnerable functions into training, validation, and testing sets with ratios of 80:10:10, respectively. This data split followed the common practice of evaluating \linevul{} in the literature (e.g.,~\cite{steenhoek2023empirical,croft2023data}).
Latent vulnerable functions, if used for training the model in each round, would be extracted from the original vulnerable functions appearing in only the training set of the round.
If any latent SVs were also flagged as originally non-vulnerable, we would consider them as (potentially) vulnerable.
We did not use latent SVs for validation and testing because of their noise.
We adopted the early stopping strategy~\mbox{\cite{hoang2019deepjit}}; we stopped training if the validation performance did not improve in the last five epochs in a round. We used the optimal model to record the measures in section~\ref{subsubsec:evaluation_measures} on the testing set.
To increase the stability of results, we ran \linevul{} 10 times in each round.
Moreover, we removed all the duplicate training entries/functions in the validation and testing sets of each round to ensure no data leakage.
We also did not use class rebalancing techniques like random over/under-sampling because they would change (training) class distributions and affect SV predictive performance, making it hard to focus on evaluating the impact of latent SVs on the performance.
Using latent SVs in conjunction with these class rebalancing techniques can be explored in the future.

\begin{table*}[t]
\fontsize{8.5}{9.5}\selectfont
  \centering
  \caption{Average and best results of function-level and line-level SV predictions using \linevul{} on the original functions in the \textit{testing} sets, \textit{with} and \textit{without} using latent vulnerable functions. \textbf{Notes}: The best values of the models among the 10 runs are in parentheses. The best row-wise values are in gray. For MFR and Effort@20\%Recall: the lower is the better.}
    \begin{tabular}{lll|c|cccc}
    \hline
    \multirowcell{3}[0ex][l]{\textbf{Dataset}} & \multirowcell{3}[0ex][l]{\textbf{Research}\\\textbf{question (RQ)}} & \multirowcell{3}[0ex][l]{\textbf{Evaluation}\\ \textbf{measure}} & \multirowcell{3}[0ex][c]{\textbf{Baseline}\\\textbf{(w/o latent SVs)}} & \multicolumn{4}{c}{\textbf{Models using original and latent SVs}} \\
    \hhline{~~~~*{4}{-}}
    & & & & \makecell{\textbf{W/o noise}\\\textbf{-reduction}} & \textbf{LIC} &  \textbf{ST} & \textbf{CR} \\
    \hline
    \multirowcell{7}[0ex][l]{\textbf{\bigvul{}}} & \multirowcell{3}[0ex][l]{RQ2 (Function-\\level)} &
    F1-Score & 0.290 (0.328) & 0.350 (0.372) & 0.339 (0.359) & 0.321 (0.332) & \cellcolor[HTML]{C0C0C0} \textbf{0.361 (0.383)} \\
    & & Precision & \cellcolor[HTML]{C0C0C0} \textbf{0.504 (0.542)} & 0.385 (0.396) & 0.444 (0.458) & 0.374 (0.410) & 0.444 (0.486) \\
    & & Recall & 0.204 (0.239) & \cellcolor[HTML]{C0C0C0} \textbf{0.321 (0.362)} & 0.275 (0.299) & 0.282 (0.301) & 0.305 (0.344) \\
    \hhline{~*{7}{-}}
    & \multirowcell{3}[0ex][l]{RQ3 (Line-\\level)} & Top-10 Accuracy & 0.802 (0.820) & \cellcolor[HTML]{C0C0C0} \textbf{0.830 (0.850)} & 0.824 (0.870) & 0.824 (0.850) & 0.824 (0.850) \\
    & & MFR & 5.226 (5.050) & 5.006 (4.830) & 5.154 (4.830) & 5.388 (4.880) & \cellcolor[HTML]{C0C0C0} \textbf{4.936 (4.450)} \\
    & & Effort@20\%Recall & 0.028 (0.027) & 0.024 (0.022) & 0.024 (0.023) & 0.024 (0.022) & \cellcolor[HTML]{C0C0C0} \textbf{0.022 (0.021)} \\
    & & Recall@1\%LOC & 0.060 (0.069) & 0.092 (0.099) & 0.092 (0.099) & 0.095 (0.109) & \cellcolor[HTML]{C0C0C0} \textbf{0.100 (0.114)} \\
    \hline
    \multirowcell{7}[0ex][l]{\textbf{\devign{}}} & \multirowcell{3}[0ex][l]{RQ2 (Function-\\level)} & F1-Score & 0.531 (0.547) & 0.613 (0.627) & 0.606 (0.613) & 0.599 (0.604) & \cellcolor[HTML]{C0C0C0} \textbf{0.617 (0.629)} \\
    & & Precision & \cellcolor[HTML]{C0C0C0} \textbf{0.637 (0.684)} & 0.489 (0.506) & 0.502 (0.553) & 0.495 (0.509) & 0.504 (0.543) \\
    & & Recall & 0.489 (0.482) & \cellcolor[HTML]{C0C0C0} \textbf{0.825 (0.909)} & 0.775 (0.881) & 0.758 (0.784) & 0.803 (0.894) \\
    \hhline{~*{7}{-}}
    & \multirowcell{3}[0ex][l]{RQ3 (Line-\\level)} & Top-10 Accuracy & 0.838 (0.850) & 0.856 (0.860) & 0.868 (0.890) & 0.862 (0.880) & \cellcolor[HTML]{C0C0C0} \textbf{0.870 (0.890)} \\
    & & MFR & 4.454 (4.380) & 4.426 (4.310) & \cellcolor[HTML]{C0C0C0} \textbf{4.126 (3.790)} & 4.438 (4.380) & 4.226 (4.080) \\
    & & Effort@20\%Recall & 0.069 (0.061) & \cellcolor[HTML]{C0C0C0} \textbf{0.041 (0.038)} & 0.043 (0.038) & 0.043 (0.038) & 0.042 (0.038) \\
    & & Recall@1\%LOC & 0.063 (0.069) & 0.072 (0.074) & 0.071 (0.074) & 0.070 (0.074) & \cellcolor[HTML]{C0C0C0} \textbf{0.073 (0.076)} \\
    \hline
    \hline
    \multirowcell{7}[0ex][l]{\textbf{Average}} & \multirowcell{3}[0ex][l]{RQ2 (Function-\\level)} & F1-Score & 0.411 (0.438) & 0.482 (0.500) & 0.473 (0.486) & 0.460 (0.468) & \cellcolor[HTML]{C0C0C0} \textbf{0.489 (0.506)} \\
    & & Precision & \cellcolor[HTML]{C0C0C0} \textbf{0.571 (0.613)} & 0.437 (0.451) & 0.473 (0.506) & 0.435 (0.460) & 0.474 (0.515) \\
    & & Recall & 0.347 (0.361) & \cellcolor[HTML]{C0C0C0} \textbf{0.573 (0.636)} & 0.525 (0.590) & 0.520 (0.543) & 0.554 (0.619) \\
    \hhline{~*{7}{-}}
    & \multirowcell{3}[0ex][l]{RQ3 (Line-\\level)} & Top-10 Accuracy & 0.820 (0.835) & 0.843 (0.855) & 0.846 (0.870) & 0.843 (0.865) & \cellcolor[HTML]{C0C0C0} \textbf{0.847 (0.870)} \\
    & & MFR & 4.840 (4.715) & 4.716 (4.570) & 4.640 (4.310) & 4.913 (4.630) & \cellcolor[HTML]{C0C0C0} \textbf{4.581 (4.265)} \\
    & & Effort@20\%Recall & 0.049 (0.044) & 0.033 (0.030) & 0.034 (0.031) & 0.034 (0.030) & \cellcolor[HTML]{C0C0C0} \textbf{0.032 (0.029)} \\
    & & Recall@1\%LOC & 0.062 (0.069) & 0.086 (0.094) & 0.082 (0.087) & 0.083 (0.092) & \cellcolor[HTML]{C0C0C0} \textbf{0.087 (0.095)} \\
    \hline
    \end{tabular}
  \label{tab:rq2_rq3_results}
\end{table*}

\subsubsection{\textbf{Evaluation measures}}\label{subsubsec:evaluation_measures}
We present the evaluation measures for function-level (RQ2) and line-level (RQ3) SV prediction. The measures in RQ3 were calculated based on the true-positive functions predicted by \linevul{} from RQ2.

\noindent \textbf{Function-level SV prediction (RQ2)}. 
We employed \textit{F1-Score}, \textit{Precision}, and \textit{Recall} to assess the performance of function-level SV prediction. These measures ranging from 0 to 1, with 1 indicating the best value, have been widely utilized in previous SV prediction studies (e.g.,~\cite{zhou2019devign,fu2022linevul,le2022use,steenhoek2023empirical}). F1-Score was considered for selecting optimal models because it is the harmonic mean of Precision and Recall. The reported results were the \textit{testing} performance of the optimal models, based on the highest validation F1-Score.

\noindent \textbf{Line-level SV prediction (RQ3)}.
We computed \textit{Top-10 Accuracy} and \textit{Mean First Rank} (MFR) commonly used to assess model interpretability~\cite{li2021vulnerability,fu2022linevul}, i.e., the effectiveness of localizing vulnerable lines, especially the first one for developers to start inspection.
Top-10 Accuracy is the percentage of vulnerable functions with at least one vulnerable line in the top-10 ranking, indicating the line localization accuracy without inspecting too many (10+) lines.
MFR is the average value of the highest rank of all the localized vulnerable lines in each function.
A low MFR value means a high accuracy of localizing the first vulnerable line for developers to inspect.

We also used \textit{Effort@20\%Recall} and \textit{Recall@1\%LOC}~\cite{fu2022linevul}. The former quantifies the LOC required to find 20\% of the actual vulnerable lines, indicating lower effort with a smaller value. The latter measures the proportion of correctly located vulnerable lines within the top 1\% LOC, meaning more SVs found using the same effort.

\subsection{RQ2 Results: Function-Level SV Prediction with Latent SVs}
\label{subsec:rq2_results}

Using latent SVs together with original SVs produced up to 24.5\% better performance (F1-Score) of function-level SV prediction than \textit{without} using latent SVs (see \tab~\ref{tab:rq2_rq3_results}).
The models with latent SVs also increased Recall by 34.8-68.7\%, while decreasing Precision by 11.9-25.8\% of the original models, probably because the datasets with latent SVs were less skewed toward the non-vulnerable class.
The better F1-Score and Recall values were statistically significant, based on the one-sided Wilcoxon signed-rank tests~\cite{wilcoxon1992individual} with $p$-values $<$ 0.01 and non-negligible effect sizes;\footnote{Effect size $(r) = Z / \sqrt{N}$, where $Z$ is the $Z$-score value of the test and $N$ is sample size; $r \geq 0.1$ means non-negligible~\cite{tomczak2014need}.}  full details are at~\cite{reproduction_package_msr2024_latent_sv}. 
These findings imply that latent SVs can increase the coverage of SV patterns, improving the overall SV predictive performance.
Moreover, \tab~\ref{tab:latent_data_analysis} shows that latent vulnerable functions add missing vulnerable functions and correct functions originally labeled as non-vulnerable in SV datasets, which are two potential factors contributing to the performance improvements of models with latent SVs.
We found that the improvements were contributed by both factors, yet more by the former (i.e., 0.332 F1 for \bigvul{} and 0.609 F1 for \devign{}) than by the latter (i.e., 0.305 F1 for \bigvul{} and 0.593 F1 for \devign{}).
In the case when all the (non-vulnerable) functions were to be included, the baseline F1-score could drop up to 38\%.
The presented results encourage the use of latent SVs, particularly to improve the performance of function-level SV prediction.

The noise-aware models performed slightly better or on par (within 5\%) compared to the noise-unaware models.
Despite being able to remove noisy latent SVs, LIC, ST, and CR discarded a large portion of latent functions in the training sets, i.e., 49\%, 39\%, and 31\% on average, respectively.
With the noise of 6\% in RQ1, these percentages highlight the challenge of automatically reducing the noise without sacrificing a large number of true-positive latent SVs, even with the customized/contemporary noise-reduction techniques.
As mentioned above, the improved performance mainly came from many newly added vulnerable functions. Discarding many of these functions would likely reduce the performance even when the noise is correctly removed.
This motivates future research to develop better noise-reduction techniques for latent SVs.

Given the observed benefits of latent SVs, we also explored the ability to automatically identify latent vulnerable functions at scale using \textit{only} trained SV prediction models (i.e., without respective original vulnerable functions in the training sets).\footnote{We did not evaluate this ability on the line level as it is very challenging to extract exact vulnerable lines in latent vulnerable functions from the lines in original functions.}
Note that without original SVs, these latent SVs could not be detected based on SZZ outputs.
When using the best models incorporating latent SVs in \tab~\ref{tab:rq2_rq3_results}, we could identify latent vulnerable functions extracted from the \textit{testing} original vulnerable functions in \bigvul{} and \devign{} with Recall of 0.377 and 0.790, respectively.\footnote{The low noise level of latent SVs would not significantly affect these values.}
Such Recall values were on par with those (0.321 for \bigvul{} and 0.825 for \devign{}) of original vulnerable functions, and around 2.5$\times$ better than those (0.155 for \bigvul{} and 0.329 for \devign{}) of baseline models without latent SVs.
The lower value of \bigvul{} was again likely due to the more diverse patterns of latent SVs in this dataset.
The results suggest that models incorporating latent SVs have the potential to identify latent SVs, even when original SVs/VFCs are not available.

\begin{tcolorbox}
\textbf{RQ2 Summary}.
Models using latent SVs significantly improve the function-level performance of baselines by up to 24.5\% in F1 and 69\% in Recall, mainly because of many newly added vulnerable functions. CR-based noise-aware models raise the performance by up to 5\%, yet discard many (31\%) of true-positive functions.
The trained models with latent SVs can also be used for detecting latent SVs with up to 80\% Recall, enlarging the current datasets.

\end{tcolorbox}

\subsection{RQ3 Results: Line-Level SV Prediction with Latent SVs}
\label{subsec:rq3_results}

Overall, incorporating latent SVs into the baseline models improved line-level SV prediction in all the metrics (see \tab~\ref{tab:rq2_rq3_results}).
The improvements were up to 3.8\%, 5.5\%, 21.4\%, and 66.7\% for Top-10 Accuracy, MFR, Effort@20\%Recall, and Recall@1\%LOC, respectively.
The latent vulnerable functions modestly improved the accuracy of localizing vulnerable lines (Top-10 Accuracy and MFR), enhancing model interpretability. Meanwhile, these functions significantly increased the cost-effectiveness of inspecting the detected vulnerable lines (Effort@20\%Recall and Recall@1\%LOC).
Latent SVs reduced the rank of vulnerable lines, by 6.1\% on average, which benefits Effort@20\%Recall and Recall@1\%LOC, but not necessarily the highest-ranked line in each original vulnerable function that directly contributes to Top-10 Accuracy and MFR.
One potential explanation for this is the missing vulnerable lines in latent SVs.
In practice, latent SVs may not always have all the vulnerable lines in their original functions; e.g., the line \code{avio\_r8(pb)} in the original vulnerable function \code{asf\_read\_marker} no longer existed in the latent vulnerable function in the intermediate commit 2 (\textit{c1fea23}) in \fig~\ref{fig:vszz_example}. 
Consequently, the top-ranked vulnerable lines in the original functions may not exist in the respective latent function(s). In contrast, many of the other vulnerable lines, not only the first/top one, still appeared in the latent vulnerable functions.
The line-level improvements over the baseline models were statistically significant, based on the one-sided Wilcoxon signed-rank tests~\mbox{\cite{wilcoxon1992individual}} with $p$-values $<$ 0.01 and non-negligible effect sizes; full details are at~\cite{reproduction_package_msr2024_latent_sv}.
Similar to RQ2, newly added (latent) vulnerable functions contributed more than corrected cases to the line-level improvements, and including all (non-vulnerable) functions could significantly reduce the baseline line-level measures by up to 23\%.

CR was again the most effective noise-reduction technique for line-level SV prediction. CR obtained better average values for all the line-level measures than the noise-unaware models. However, the differences between them were rather small (1-3\%). A possible reason is that many of the vulnerable lines needed to further improve the line-level measures were actually in the significant portion discarded (31\%) of latent SVs, as shown in section~\ref{subsec:rq2_results}.

\begin{tcolorbox}
\textbf{RQ3 Summary}.
Models using latent SVs improve the line-level SV prediction, up to 5.5\% for the accuracy and up to 66.7\% for the cost-effectiveness of vulnerable line localization. CR-based noise reduction performs the best and further enhances the results by up to 3\%.

\end{tcolorbox}

\section{Discussion}
\label{sec:discussion}

\subsection{Latent SVs and Beyond}

\noindent \textbf{Potential} SV Data Augmentation with Latent SVs.
We highlight the potential of latent SVs for data augmentation of vulnerable code.
Data augmentation increases the size of data by creating new samples based on existing ones~\cite{bayer2022survey}.
A key challenge of applying data augmentation to code is to preserve/guarantee the labels of generated samples~\cite{zhuo2023data}.
In the context of vulnerable code, synthetic code generation~\cite{rabin2021generalizability}, like swapping code statements, may not preserve SV semantics in the original code, making it extremely challenging to guarantee the SV-proneness/label of the generated code.
Theoretically, latent SVs provide a natural and readily available way to augment real-world vulnerable code. This is possible because with respect to the original vulnerable functions, these latent functions likely share the same SV label (vulnerable lines unchanged), yet with different code structures due to real code changes between introducing and fixing commits.
Practically, the RQ results show that latent SVs are abundant (4$\times$ the original SV data size), have low-level noise ($\approx$6\% with the current SZZ-based detector), and can significantly improve the performance of function-level and line-level SV predictions.
Because of these reasons, we recommend further exploration/utilization of latent SVs for SV management and analysis tasks other than SV detection~\cite{le2019automated,le2021large,duan2021automated,le2022survey,le2022towards,fu2022vulrepair}.

\noindent \textbf{Use of Latent SVs for Low-Resource Projects}.
We observed that \bigvul{} had fewer latent SVs than \devign{}; however, the function-level and line-level improvements over the baselines of \bigvul{} were generally similar or even higher than those of \devign{} (see \tab~\ref{tab:rq2_rq3_results}).
This finding suggests that latent SVs can be particularly useful for enhancing the SV predictive performance for projects with a limited number of original SVs.
Fewer original SVs in a project typically mean a lower variety of SV patterns for a model to learn. In such cases, with their large size, latent SVs would likely provide more diverse SV patterns than those offered by the original SVs.
This finding is further reinforced as we found that on average, using only 60\% of the original data together with all of their respective latent SVs could already produce a similar performance to that when all of the original vulnerable functions were used.\footnote{We compared models using 10--90\% (10\% increment) of original SVs + their latent SVs with models using all original SVs without latent SVs. More details are given at~\cite{reproduction_package_msr2024_latent_sv}.}
Overall, latent SVs provide a promising and convenient way of improving data-driven SV prediction for low-resource projects, without requiring additional efforts of manual data labeling.

\noindent \textbf{Non-SZZ Based Identification of Latent SVs}.
SZZ-based latent SVs are of decent quality, but we can only extract them when there are original SVs fixed in VFCs. In practice, many SVs have been silently fixed and not publicly reported~\cite{nguyen2022hermes,sawadogo2022sspcatcher}.
We cannot use SZZ, and thus need non-SZZ techniques, like using an SV prediction model itself, to identify latent SVs when VFCs are missing.
Our results show that the best models trained on original SVs and their latent SVs can detect some of these latent SVs, up to nearly 80\% Recall. Yet, for low-resource projects with few original SVs, Recall can be much lower, i.e., only $\approx$40\% for \bigvul{}. Automatic identification of (missing) VFCs like D2A~\cite{zheng2021d2a} is another potential solution to enhancing the detection of latent SVs, yet such approaches still require refinement to reduce noise/false positives~\cite{croft2023data}.

\subsection{Threats to Validity}\label{subsec:threats_to_validity}

The first threat is the completeness of latent SVs. We tackled it by adhering to the best practices and employing the state-of-the-art \vszz{} algorithm to identify latent vulnerable functions from original SVs. However, achieving 100\% completeness is impractical as some of the original SVs/VFCs were absent from the studied datasets.
Ideally, test cases are needed to confirm the existence of SVs and increase the coverage of latent SVs. However, existing SV datasets do not include such test cases, and it is also extremely tedious to manually create test cases~\cite{bui2022vul4j}.

The potential subjectivity in the manual labeling process can also be a concern. To mitigate this, two authors, possessing a combined experience of 10+ years in the field, independently performed manual validation on significant samples. Moreover, we enhanced the reliability of the validation process by cross-referencing the information (vulnerable lines) from the original SVs.

Generalizability and reliability are also potential concerns. To address generalizability, we utilized two widely used SV datasets comprising 300+ real-world C/C++ projects.
We also only investigated one SV prediction model, i.e., \linevul{}, but it is the state-of-the-art and commonly used for function-level and line-level SV predictions. Thus, our results can directly inform the practice of data collection and usage for follow-up research in this area.
We made our data/code available at~\cite{reproduction_package_msr2024_latent_sv} for others to extend our research to other programming languages and SV prediction models.
For addressing reliability, we employed the Wilcoxon signed rank test and its effect size, to rigorously assess the significance of findings.

\section{Conclusion}
\label{sec:conclusions}

We highlighted the value of latent SVs to improve SV datasets and SV prediction models. We customized the state-of-the-art \vszz{} algorithm to identify 100k+ latent vulnerable functions in the \bigvul{} and \devign{} datasets. Then, we explored the quality and prevalence of these latent SVs with respect to the original vulnerable and non-vulnerable functions.
We found that 90\% of these latent SVs (with 6\% noise) were missing in the datasets and they could increase the SV data size by 4$\times$ on average and correct 5k mislabeled functions.
After incorporating such latent SVs into the state-of-the-art \linevul{} SV prediction model, the performance (F1-Score) of function-level SV prediction improved by up to 24.5\%, while the effectiveness of line-level SV prediction was boosted by up to 67\%.
These results encourage further use of latent SVs as a cost-effective and powerful technique for advancing downstream data-driven SV management tasks, especially in low-resource scenarios.

\section{Data Availability}
\label{sec:data_avail}
The data and code of this study are available at~\cite{reproduction_package_msr2024_latent_sv}.

\section*{Acknowledgments}
The work has been supported by the Cyber Security Research Centre Limited whose activities are partially funded by the Australian Government's Cooperative Research Centres Program.

\balance

\bibliographystyle{ACM-Reference-Format}
\bibliography{reference}

\end{document}